# Neutron reflectometry with registration of secondary radiation

V.D. Zhaketov[1], A.V. Petrenko[1], Yu.M. Gledenov[1], Yu.N. Kopatch[1],
N.A. Gundorin[1], Yu.V. Nikitenko[1,*], V.L. Aksenov[1]

[1]*Joint Institute for Nuclear Research, Dubna, Russia*
*\*nikiten@nf.jinr.ru*

Neutron reflectometry is a method for measuring the spatial profile of the neutron interaction potential with the medium. The interaction potential is the sum of neutron interaction potentials with separate isotopes of the medium. To determine the neutron interaction potential with the structure units, secondary radiation is simultaneously registered with neutrons. Channels for registering secondary radiation of charged particles, gamma rays and neutrons, having experienced spin flip, have been developed on the REMUR spectrometer of the IBR-2 reactor in Dubna (Russia). The necessity of registering secondary radiation in neutron reflectometry is justified, a method developed for measuring secondary radiation is described and the results of testing channels for registering secondary radiation on the REMUR spectrometer are presented in the paper.

**Introduction.** Currently, it is relevant to study the phenomena of proximity occurring at the interface between two media, namely, superconductor and ferromagnetic [1-3]. Due to mutual impact of ferromagnetic and superconductor, caused by the finite values of the coherent wave lengths of phenomena, there is significant modification of magnetic phenomena and superconducting properties. This becomes obvious, in particular, in the change of the spatial distribution of magnetization. In this case, it is important to establish the correspondence of the magnetic spatial profile (spatial dependence of magnetization) with the nuclear spatial profiles of elements(isotopes) of the contacting layers. For determining the spatial magnetic profile, standard reflectometry of polarized neutrons is used, which measures the energy of the nuclear-magnetic interaction (hereinafter, the interaction potential) of a neutron with the medium [4].

At the interface between two media, the interaction potential is the sum of interaction potentials of elements penetrating each other. In this regard, standard neutron reflectometry does not allow one to identify to what kind of elements the changes of the interaction potential and, in particular, of the magnetic profile, are related. For defining the potential profile of the neutron interaction with separate elements, it is necessary to register secondary radiation from isotopes of elements [5]. The type of radiation and the radiation energy are the signs identifying isotopes of elements. Charged particles, gamma rays and nuclear fission fragments are considered as secondary radiation. In a broader interpretation, secondary radiation should include neutrons, incoherently scattered at nuclei, neutrons inelastically scattered by atoms and the medium, as well as diffusely scattered neutrons at interfaces and inhomogeneities in the layers of the structure. The signs of such secondary radiation are the angular distribution and the transfer energy of scattered neutrons. The particular secondary radiation includes neutrons having experienced a coherent spin flip in a noncollinear magnetic structure. In this case, at neutron propagation, neutrons are "absorbed" in the initial spin state and, correspondingly, neutrons occur in final spin state. For determining the spatial distribution of elements and increasing the sensitivity of measurements, neutron standing wave regime are used [6,7].

In [8,9], the registration of secondary gamma radiation from the structures containing a 5 nm-thick gadolinium layer was reported. In [10], secondary radiation was registered in the form of charged particles emitted by the $^6$Li isotope after neutron capture by them. In [11, 12], the REMUR neutron reflectometer (Dubna, Russia) was described, in which charged particles and gamma rays were registered together with neutrons. In this paper, justification of the necessity for registration of secondary radiation has been performed. Model calculations of the neutron absorption coefficient have been carried out for the case of bilayers and periodic structures. Developed on the REMUR spectrometer, the channels for registering secondary radiation, namely, charged particles, gamma rays and spin-flipped neutrons are described. The results of channels testing are presented. Prospects for the further development of the current neutron reflectometry method are outlined.

**Basic relations.** For the range of the neutron wave vector $k < \pi/d$, where $d$ is the distance between atoms of the medium, neutron propagation in the medium is described by the interaction potential of the neutron with the medium [13]:

$$U(z) = V - iW = \alpha \Sigma_g N_g b_g + \boldsymbol{\mu} \boldsymbol{B} - i[\beta(\Sigma_g N_g \Sigma_p \sigma_{gp} + N_n \sigma_n + N_m \sigma_m + N_{nm} \sigma_{nm}) + \gamma_{ne} \sigma_{ne}] \quad (1)$$

where $N_g(z)$ is density of the g-type nuclei, $b_g$ is the amplitude of elastic coherent neutron scattering by the g-type nucleus, $\sigma_{g,p}$ is the interaction cross section with the *g*-type nucleus to form *p*-radiation with the energy $E_p$, $\sigma_n$, $\sigma_m$, $\sigma_{nm}$ and $N_n$, $N_m$, $N_{nm}$ are the neutron scattering cross section and the density of nuclear, magnetic and nuclear magnetic inhomogeneities of the medium, $\sigma_{ne}$ is the inelastic neutron scattering cross section by the medium, $\gamma_{ne}$ is the dimension factor, $\boldsymbol{B}$ is the magnetic field induction vector, $\alpha = 2\pi\hbar^2/m$, $\beta = \hbar v/2$, $\hbar$ is the Planck constant, $\boldsymbol{\mu}$ is the neutron magnetic moment, $m$ is the neutron mass, $v$ is the neutron velocity. The neutron transport through a layered structure follows the reflection coefficient $R(Q)$, the transmission coefficient $T(Q)$ and the neutron absorption coefficient $M(Q)$. For coefficients, the flux balance equation for the initial "*i*" and final "*j*" spin states of the neutron is

$$\Sigma_j \left( R^{ij}(Q) + T^{ij}(Q) + M^{ij}(Q) \right) = 1 \quad (2)$$

The partial absorption coefficient is the ratio of the absorbed neutron flux $J_g^j(E_p)$, corresponding to the secondary radiation with the energy $E_p$, to the initial neutron flux $J_0^i$

$$M_{g,p}^{ij} = \frac{J_g^j(k_z^j, E_p)}{J_0^i(k_{0z})} = \int \frac{[J^j(z,k_z^j) = n^j v_z^j] N_g(z) \sigma_{g,p}(k_z^j) dz}{J_0^i} \quad (3)$$

where $J^j(z, k_z^j) = n^j(z, k_z^j) v_z^j$ is neutron flux in medium, $n^j(z, k_z^j)$ is neutron probability density (hereinafter, just neutron density), $v_z^j$ is $z$ – component of neutron velocity, $k_z = \sqrt{k_{0z}^2 - (k_v^2 - ik_w^2) - \boldsymbol{\mu}\boldsymbol{B}}$ and $k_{0z} = k_0 \sin\theta$ are $z$ - components of the neutron wave vector in the medium and vacuum, correspodently, $k_v = \frac{\sqrt{2mV}}{\hbar}$ and $k_w = \frac{\sqrt{2mW}}{\hbar}$ are critical wave vectors, connected with real $V$ and imaginary $W$ parts of potential, correspondently, $\boldsymbol{\mu}$ is neutron magnetic moment, $\theta$ is initial glancing angle.

Further, it will be useful to represent $M_{g,p}^{ij}$ in terms of other physical quantities. Thus, $M_{g,p}^{ij}$ can be expressed in terms of imaginary part of the partial interaction potential:

$$W_{g,p} = \hbar v_z N_g \sigma_{g,p}/2; \quad M_{g,p}^{ij} = \frac{2m}{\hbar^2} \int \frac{W_{g,p} n^j}{n_0^i k_{0z}^i} dz \qquad (4)$$

A different view using wave functions:

$$M_{g,p}^{ij} = \int |y^{ij}|^2 \frac{k_{g,p}^2}{k_{0z}^i} dz \qquad (5)$$

where $y^{ij}(z, k_z) = \frac{\psi^j}{\psi_0^i}$, $k_{g,p}^2 = \frac{2m}{\hbar^2} W_{g,p} = N_g \sigma_{g,p} k_z^j$.

As it follows from (5), $M_{g,p}^{ij}$ is determined by the coefficient of transformation of the neutron density $|y^{ij}|^2$. In the contrast standing wave regime $|y^{ij}|^2$ is in the range from zero to four and in the enhanced standing wave regime $|y^{ij}|^2$ can reach $4 \cdot 10^4$ for real structures, while due to increased neutron density with spin flip to the third power in respect to increased neutron density without spin flip, $|y^{ij}|^2 = (|y^{ii}|^2)^3$, where $i \neq j$ [14-16].

The neutron absorption coefficient corresponding to unregistered secondary radiations is determined from the relation

$$M_{res}^i = 1 - R^i + T^i - \Sigma_p M_{1p}^i - \Sigma_s M_{2s}^i \qquad (6)$$

where $M_1$ and $M_2$ are absorption coefficients of two contacting layers.

In the case of two isotopes at the interface, in order to solve (6) it is sufficient to identify the neutron absorption coefficient in one of the isotopes.

We now turn to the issue of the neutron wave field. Standard reflectometry, measuring the reflection and transmission coefficients, "operates" in the traveling neutron wave regime. The neutron density in this case decays deep into the structure at an exponential rate. Due to the phase change of the neutron wave at passing a separate layer of the structure on $\pi$, the layer thickness is defined from the relation $d = \pi \left(\frac{1}{k_{z2}} - \frac{1}{k_{z1}}\right)$, where $k_{z1}$ and $k_{z2}$ are the values of the wave vector at which the neutron reflection intensity has local maxima.

The imaginary part of potential at registering the secondary radiation allows one to identify isotopes of elements, but its impact on absorption of neutrons in the traveling wave mode is very little, especially in the important case of thin layers of angstrom thickness. To increase the neutron absorption in the investigated layer and to determine its spatial position, a neutron contrast standing wave regime (neutron density changes from zero to four) is used. In order to create the contrast a standing wave regime, there must be a neutron reflector in the structure.

The neutron flux in vacuum in front of the reflector is a periodic function:

$$j(z, k) \equiv v_z \cdot |\psi(z, k) = \psi_d(z, k) + \psi_b(z, k)|^2 = v \cdot n_0(0, k)[(1 - |r_r| \exp(-2k_I z))^2 + 4|r_r| \exp(-2k_I z) \cos^2(k_r z + \varphi_r/2)] \qquad (7)$$

where $\psi_d$ and $\psi_b$ are wave functions in direct and back directions, correspondingly, $r_r$ is the reflection amplitude from the reflector, $\varphi_r$ is the phase of the neutron reflection amplitude, $k_R$ and $k_I$ are the real and imaginary parts of the wave vector in the layer, correspondingly. At $|r_R| \approx 1$, $k_I = 0$ and in case $k_R z_{max} + \varphi_r/2 = n\pi$, the density increases relatively density of incident neutrons four times ($n = 4n_0$). With the increase of the flux $j = n v_z$, the neutron absorption coefficient in the layer positioned at $z_{max}$ will increase. As a result, by the value of $k_R$ at which the absorption coefficient has a maximum, the layer position is defined.

At interference of waves in the area of structure between two interfaces in phase-shifting layer [6], a resonant regime of neutron standing waves occurs. For the neutron flux density in this case we have:

$$j(z, k_z) = v_z \cdot n_{sw} \cdot \left\{ \eta_{esw} = \left[|t_a|^2 |t_{vp}|^2 / |1 - r_a \cdot r_{p+r}|^2\right] / \left[(1 - |r_{pr} \times r_{pa}| \times \exp(-2k_{z,I}L))^2 + 4|r_{pa} \times r_{pr}| \times \exp(-2k_{z,I}L) \sin^2\left(k_R L + \frac{\varphi_{fa}}{2} + \frac{\varphi_{fr}}{2}\right)\right] \right\} \qquad (8)$$

where $n_{sw}$ is the density of standing waves and $\eta_{esw}$ is the density transformation factor resulting from wave interference, $t_a$ is amplitude of the transmitted wave of the amplifying layer [6], $t_{vp}$ is amplitude of the transmitted wave of the "vacuum - phase-shifting layer" interface, $r_{p+r}$ is reflection amplitude of structure "phase-shifting layer + reflector", $r_a$ is amplitude of the reflected wave from the amplifying layer, $r_{pa}$, $\varphi_{pa}$ are, correspondingly, amplitude and phase of the amplitude of the reflected wave from the structure of the "phase-shifting layer interface + amplifying layer ", $r_{pr}$, $\varphi_{pr}$ - are, correspondingly, amplitude and phase of the amplitude of the reflected wave from the structure of the "phase-shifting layer interface + reflector", $L$ is thickness of the phase-shifting layer.

In case of the resonance $Lk_z + \varphi_{pa}/2 + \varphi_{pr}/2 = m\pi$, the density transformation factor has the form

$$\eta_{esw} = \frac{n}{n_0} = \left[|t_a|^2 |t_{vp}|^2 / |1 - r_a r_{p+r}|^2\right] / \left[1 - |r_{pa} \times r_{pr}| \times \exp(-2k_{z,I}L)\right]^2 \qquad (9)$$

At $r_{fa} \approx 1$, $r_{fr} \approx 1$ and $k_{z,I} \ll 1/L$, $\eta_{esw} > 1$ is performed. Estimates show, that in the case of slightly absorbing nuclear layers, $\eta_{esw}$ can reach about $4 \cdot 10^4$. In the case of a magnetic structure, the gain value of the neutron reflection coefficient with spin flip under certain conditions increases quadratically and can reach $1.6 \cdot 10^9$ [14].

**Model calculations.**

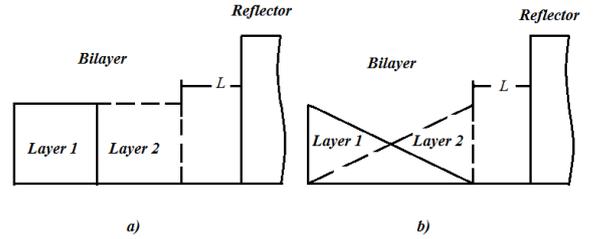

**Fig.1.** Spatial dependence of the potential of neutron interaction with a structure: a) rectangular form of layer potentials b) triangular form of layer potentials. First layer is shown in solid line, second layer – in dashed line.

The structural model which has been used for calculations is shown in Figure 1. The bilayer consisting of two layers 1 and 2 is positioned at distance $L$ from the neutron reflector. Two types of potentials are shown. In the first case, the potentials of the layers have a shape of rectangle adjacent to each other. In the second case, the potentials are in the shape of triangles overlapping on the thickness of the bilayer. As a result, the thickness of the interface varies from zero (fig. 1a) to the thickness of the bilayer (fig. 1b).

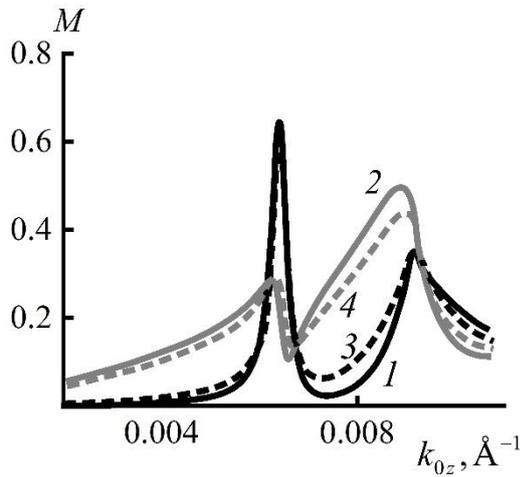

**Fig.2.** Dependence of the neutron absorption coefficient $M(k_{0z})$ in the first (1,3) and the second (2,4) layers for structure Layer1(25nm)/Layer2(25nm)/Vacuum(20nm)/Reflector with rectangular (1, 2) and triangular (3,4) layer potentials. Reflector critical wave vector is $k_v=0.009$Å$^{-1}$.

Figure 2 shows the wavelength dependences $M(k_{0z})$ in two separate layers of a bilayer 25 nm thick with rectangular and triangular type of potentials in the case of the copper reflector and the distance from the reflector to the bilayer $L=20$ nm. As it is shown, the dependences of the absorption coefficient in two layers are substantially different, which reflects their spatial position in the bilayer. The form of layer potentials (the interface form) is reflected in the max value.

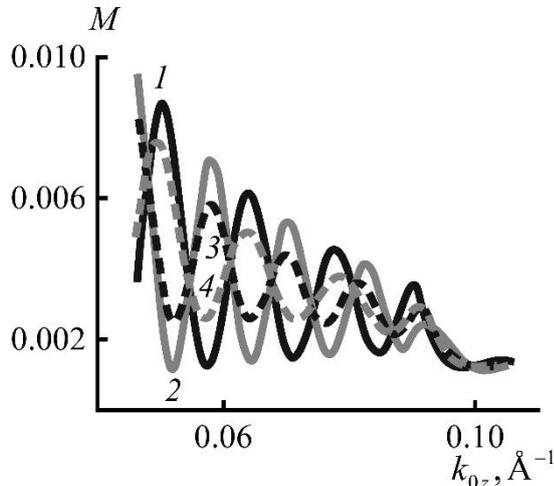

**Fig.3.** Dependence of the neutron absorption coefficient $M(k_{0z})$ in the first (1,3) and the second (2,4) layers for structure Layer1(2.5nm)/Layer2(2.5nm)/Vacuum(20nm)/Reflector with rectangular (1, 2) and triangular (3,4) layer potentials. Reflector critical wave vector is $k_v=0.09$Å$^{-1}$.

Figure 3 shows dependences similar to Figure 2, but for the case of a super-mirror reflector for which the critical wave vector $k_v$ is 10 times higher than the value for a copper reflector ($k_v=0.009$Å$^{-1}$). Obviously, in such case, as a result of decrease in the maximum value of the period of standing waves, the spatial resolution is 10 times higher. When this occurs, significant changes in dependences are observed for a thinner bilayer with a thickness of 5 nm. Now we consider the case of the magnetically noncollinear structure and polarized neutrons.

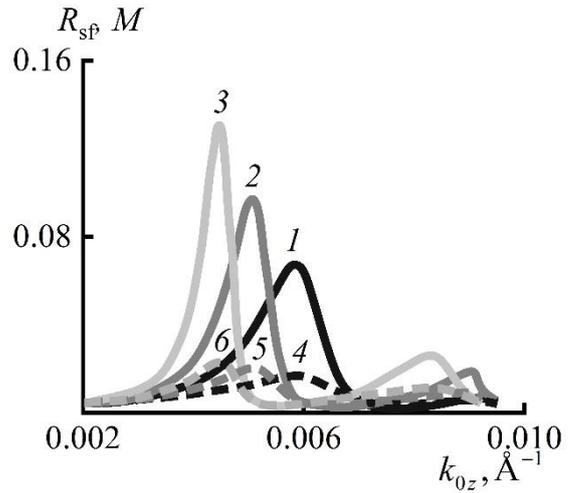

**Fig.4.** Dependences of the neutron absorption coefficient $M$ (upper dependences 1–3) and the neutron reflection coefficient with spin flip $R_{sf}$ (lower dependences 4–6) for a layer 20 nm thick with magnetization of 1 kG perpendicular to the quantization axis positioned at a distance $L=0$ (1 and 4), 15 (2 and 5) and 30 nm (3 and 6) from the copper reflector with $k_v=0.009$ Å$^{-1}$.

Figure 4 shows the dependences of the reflection coefficient with spin flip (upper curves 1,2,3) and the dependences of the absorption coefficient (lower curves 4,5,6). As it is shown, for the same values of $L$, the maxima for the upper and lower dependences coincide. This indicates that in the case of a magnetic noncollinear structure, the dependence $R_{sf}(k_{0z})$, determined by the magnetic part of the potential, as well as the dependence $M(k_{0z})$, defined by the imaginary part of the potential, can be used to identify spatial distribution of the potential.

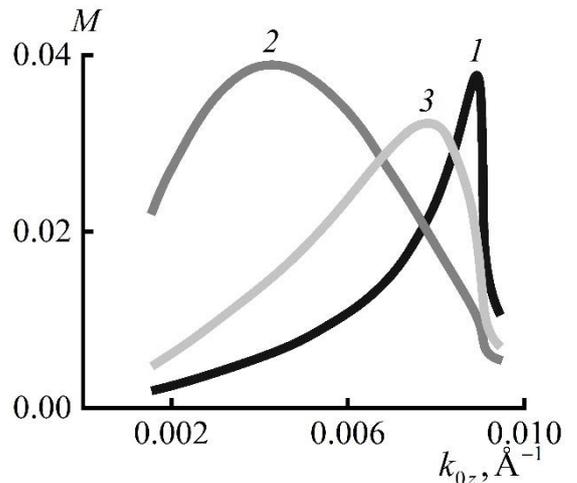

**Fig. 5.** Neutron absorption coefficient with plus $M^+(1)$ and minus $M^-(2)$ initial polarization in a layer 20 nm thick, characterized by a collinear magnetization of 10 kG. The absorption coefficient in layer without magnetization is $M(J=0)$ (3). The layer is characterized by the values $k_v=0.0053$ Å$^{-1}$ and $k_w=0.00053$ Å$^{-1}$ and is positioned at a distance $L=25$ Å from the reflector with $k_v=0.009$ Å$^{-1}$.

In the case of a collinear magnetic structure (Fig. 5), the dependence $M(k_{0z})$ is defined by the initial neutron spin state $P_0=\pm 1$. Obviously, in this case, the position of the maxima in the dependences 1 and 2 is determined by the magnetization of the layer.

We now consider the dependences of the absorption coefficient for a periodic structure. In this case, a neutron standing wave is formed by a periodic structure both in the structure and in front of it. The spatial period of a standing wave in this case is equal to the period of the periodic structure. Figure 6 shows the calculation dependence of the neutron density transformation coefficient $D$ in the model structure $50\times[Cu(3nm)/Vacuum(3nm)]$ depending on the current number of the bilayer $n$.

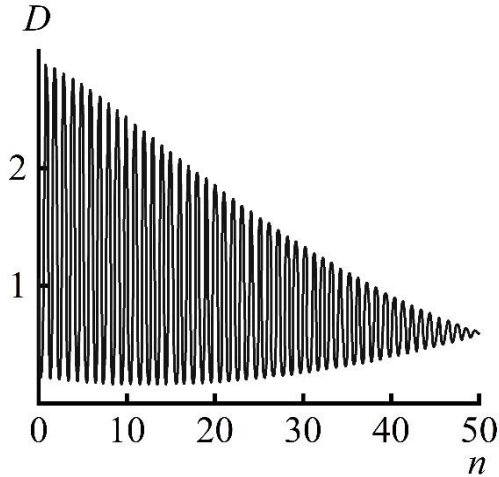

**Fig.6.** Neutron density transformation coefficient $D(n)$ in the periodic structure $50\times[Cu(3nm)/Vacuum(3nm)]$ depending on the current number of the bilayer.

The neutron density in the structure is set to maximum at the "vacuum-Cu" interface (the potential gradient is positive) and to minimum - at the "Cu-vacuum" interface (the potential gradient is negative). As it is shown, the density changes in the bilayer are at maximum in the beginning of the structure and at minimum - at the end (density of the initial neutron wave is accepted as unity). At the end of the structure, the average value of the neutron density is less than one (the input density is accepted as unity), which is related to the neutron absorption in copper layers.

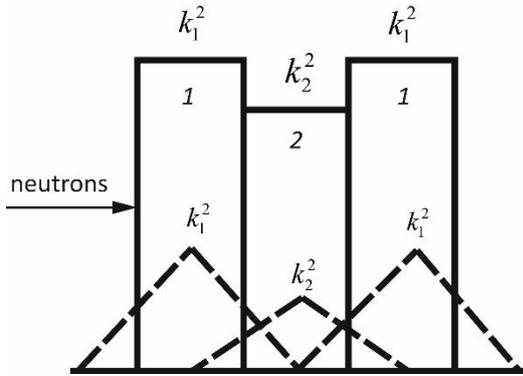

**Fig. 7.** Shape of the potentials in the periodic structure: rectangular- solid line and triangular - dashed line.

Figure 7 shows the forms of bilayer potentials used in calculations of the periodic structure. The rectangular potential has a fixed potentials $k_1^2$ and $k_2^2$ in the bilayer layers. The triangular potentials have a maxima in the middle of the one layer and equals to zero in the middle of the contacting layers.

For numerical calculations, the following potentials have been used. The rectangular potential: for the first layer $k_1^2 = 10^{-6}\times(1-i\times0.01)$ Å$^{-2}$, for the second one - $k_2^2 = 8.1\cdot10^{-5}\cdot(1-i\cdot0.001)$ Å$^{-2}$. The triangular potential: for the first layer $k_1^2 = k_{1max}^2\cdot(1 - 2|z|/d)$, where $k_{1max}^2 = 1.3\cdot10^{-6}\cdot(1-i\cdot0.01)$ Å$^{-2}$, for the second layer $k_2^2 = k_{2max}^2\cdot(1-2z/d)$, where $k_{2max}^2 = 1.1\cdot10^{-4}\cdot(1-i\cdot0.001)$ Å$^{-2}$, $d$ is thickness of the layers, the coordinate $z$ is counted from the center of the layers.

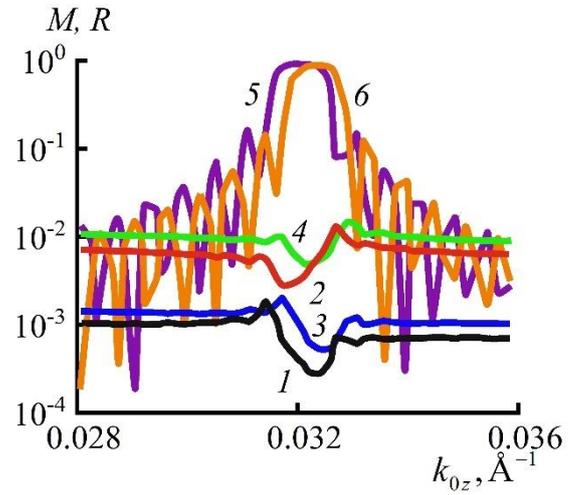

**Fig.8.** Neutron absorption coefficient in the first (1,3) and in the second layer (2,4) of the bilayer of the structure $50\cdot[k_2^2(5nm)/k_1^2(5nm)]/k_3^2$ with the rectangular layer potentials $k_1^2(z)=10^{-6}$ of the layer, $k_{1max}^2 = 1.3\cdot10^{-5}\cdot(1-i\cdot0.01)$ Å$^{-2}$, $k_{2max}^2=1.1\cdot10^{-4}\cdot(1-i\cdot0.001)$ Å$^{-2}$ (3,4). $d=5$ nm – thickness of the first and the second layer. $k_3^2 = k_{3v}^2-i\,k_{3w}^2$, $k_{3v}^2=10^{-8}$ Å$^{-2}$, $k_{3w}^2=10^{-4}\cdot k_{3v}^2$ Å$^{-2}$. Reflection coefficient for rectangular (5) and triangular (6) potentials.×(1-i·0.01) Å$^{-2}$, $k_2^2(z)=8.1\cdot10^{-5}\cdot(1-i\cdot0.001)$ Å$^{-2}$ (1,2) and with the triangular layer potentials $k_1^2 = k_{1max}^2\cdot(1 - 2|z|/d)$, $k_2^2 = k_{2max}^2\cdot(1 - 2|z|/d)$, where $k_{1max}^2$, the counting of the coordinate $z$ from the center

Figure 8 shows the neutron absorption coefficient $M(1–4)$ and the neutron reflection coefficient $R(5.6)$ for the structure $50\times[k_2^2(5nm)/k_1^2(5nm)]/k_3^2$ depending on the neutron wave vector component $k_{0z}$. The absorption coefficients are presented for the first (1,3) and the second (2,4) layers of the bilayer in the cases of rectangular (1,2) and triangular (3,4) potentials. The reflection coefficient $R$ is shown both for the rectangular (5) and the triangular (6) potentials.

As we can see, the form of the potential determines the typical range of variations of the wave vector. Indeed, all the dependences for the triangular potential are shifted towards high values of the wave vector with respect to the dependences for the rectangular potential.

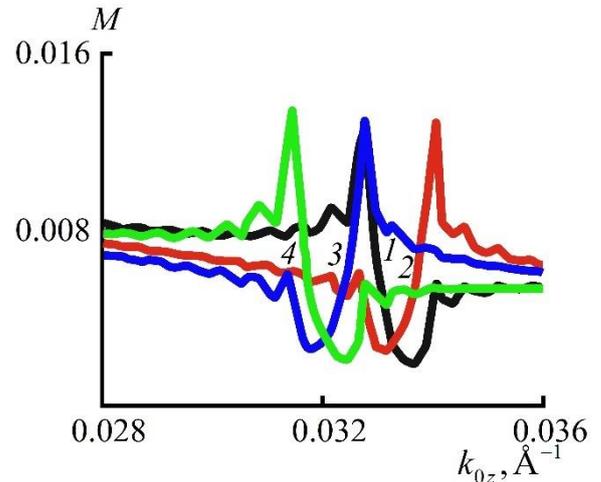

**Fig.9a.** Absorption coefficient in the first (1,3) and the second (2,4) layers of the bilayer of the structure $50\times[k_2^2(5nm)/k_1^2(5nm)]/k_3^2$ with the layer rectangular nuclear potentials and the magnetization of 28.3 kG in the second layer of the bilayer for positive (1,2) and negative (3,4) neutron polarization.

Figure 9a shows dependences of the absorption coefficient for the neutron polarization "+" and "-". At polarization "+", the potential in the second layer, which is the sum of nuclear and magnetic potentials, exceeds the nuclear potential in the first layer. With the neutron polarization "-", the potential of the second layer is zero and the potential in the first layer keeps being equal to the nuclear potential. As it is shown in Fig. 9a, the dependences for the polarizations "+" and "-" turn out to be symmetric with respect to the maximum at $k_{0z}=0.0326$ Å$^{-1}$ in the dependences of the absorption coefficient for the first layer.

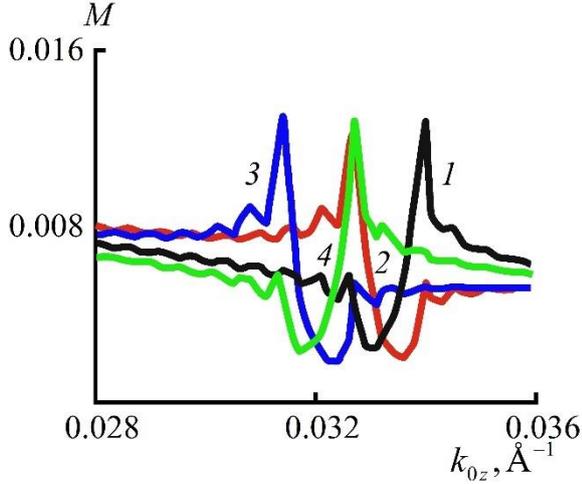

**Fig.9b.** Neutron absorption coefficient in the first (1,3) and the second (2,4) layers of the bilayer of the structure 50×[ $k_2^2$ (5nm)/ $k_1^2$(5nm)]/$k_3^2$ with the layer rectangular nuclear potential $k_1^2= k_2^2=8.1\cdot10^{-5}\cdot(1-i\cdot0.001)$ Å$^{-2}$ and the magnetization of 28.3 kG in the first layer of the bilayer with positive (1,2) and negative (3.4) polarization of falling neutrons.

Figure 9b shows dependences similar to those in Figure 9a, but in this case the magnetization is in the first layer. As it is presented, the dependencies for layers of the bilayer are reversed. Now the dependences are symmetric with respect to the maximum in dependences of the second layer, which in this case has a nuclear potential. We can note that the reflection coefficients for two polarizations are also symmetrical in regard to the same value of the wave vector, but they aren't reversed with the change of the magnetization position. Thus, the reflection coefficients do not allow one to determine the position of the magnetization. On the contrary, the absorption coefficients provide this information.

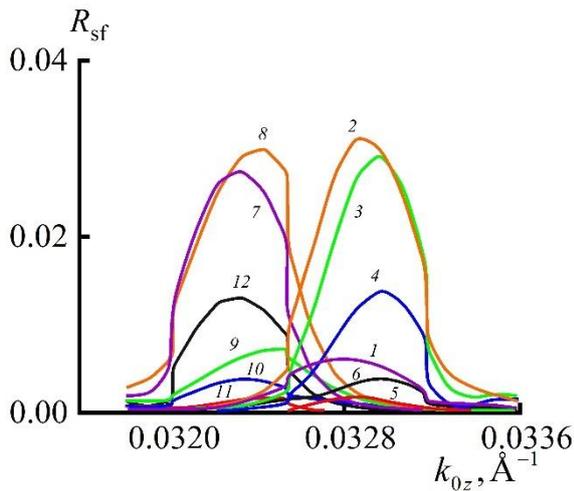

**Fig.10**. Neutron reflection coefficients with spin flip $R^{+-}$ and $R^{-+}$ for the structure 50×[ $k_2^2$ (5nm)/ $k_1^2$(5nm)]/$k_3^2$ at collinear magnetization of the first layer $J_z$=28.3 kG. The sublayer 1 nm thick with magnetization perpendicular to the quantization z-axis $J_x$=0.1$J_z$ is at the beginning (1, 7), in the middle (2, 8) and at the end (3, 9) of the second layer and at the beginning (4, 10), in the middle (5, 11) and at the end (6, 12) of the first layer of the bilayer. The dependencies 1-6 correspond to $R^{+-}$, the dependencies 7-12 - to $R^{-+}$. The nuclear rectangular potentials in both layers of the bilayer are $k_1^2= k_2^2=8.1\cdot10^{-5}\cdot(1-i\cdot0.001)$ Å$^{-2}$.

Figure 10 shows the dependences of the neutron reflection coefficient with neutron spin flip at different positions in the bilayer of the magnetic noncollinear layer 1 nm thick. For the positive initial polarization, the potential of the first layer is higher than the potential of the second one, for the negative polarization, on the contrary, the potential of the second layer is higher than the potential of the first layer. As a result, the spatial dependence of the neutron density, as well as the dependences $R^{+-}(k_{0z})$ and $R^{-+}(k_{0z})$ for two polarizations will be different. Apparently, the flux density is maximum in the middle of the second layer (2 and 8) and minimum in the middle of the first layer (5 and 11) for the both polarizations. In contrast, in other points of the bilayer, the neutron flux density is different for polarizations of different signs. Thus, the noncollinear layer in this case serves as an analyzer of the neutron flux density in the bilayer. Obviously, the use of neutrons with different polarizations allows one to localize the position of the magnetic-noncollinear layer. It must be stated that the absorption coefficients do not change at variation of the magnetic-noncollinear sublayer position, and they have the form shown in Figures 9a,b. This is due to the fact that a thin magnetic-noncollinear layer does not significantly change the distribution of the neutron flux. For the first time, M.A. Andreeva (Moscow State University, Moscow) drew our attention to the change in the spatial dependence of the neutron density in the periodic structure with variation of the neutron polarization.

**Channels for registering secondary radiation.**
Registration of secondary radiation at the grazing incidence geometry is implemented on the spectrometer REMUR [17], positioned on the channel 8 of the pulsed reactor IBR-2 in Dubna (Russia). The IBR-2 reactor operates with a pulse-repetition rate of 5 Hz, the neutron pulse width in the water moderator is 340 μs. The investigated structure on REMUR is installed at a distance of 29 m from the neutron moderator of IBR-2. The position-sensitive neutron detector is positioned at a distance of 4,9 m from the place of the sample installation. The root-mean-square deviation of the wave length registered by the neutron detector is 0.02 Å. Three channels for registering secondary radiation are designed and tested on the spectrometer: the channel of polarized neutrons, the channel of charged particles and the channel of gamma rays.

**Channel of polarized neutrons**. The functional scheme of the channel for registering polarized neutrons is presented in Fig. 11. The channel includes polarizer P, input SF-1 and output SF-2 neutron spin flippers, AP polarization analyzer and neutron detector D. The structure being investigated S is arranged between spin flippers. The polarizer consists of a super-mirror structure deposited on a glass substrate. The polarizer substrate is produced from a polished piece of glass with a size of 10cm×80cm×2.8cm. The super-mirror coating has been implemented at ILL (Grenoble, France) and is characterized by a critical neutron reflection angle of 3.5 mrad/Å (a type of super-mirror m2).

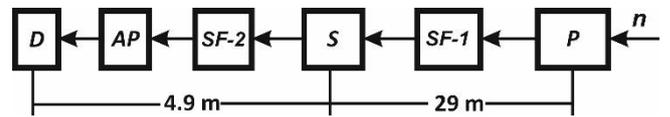
**Fig. 11.** Functional scheme of the polarized neutrons channel.

The polarization efficiency of the polarizer $P_p(\lambda)$ (curve 1) is shown in Fig.12. $P_p(\lambda)$ exceeds 0.9 in the wavelength range 1.3–3.25 Å.

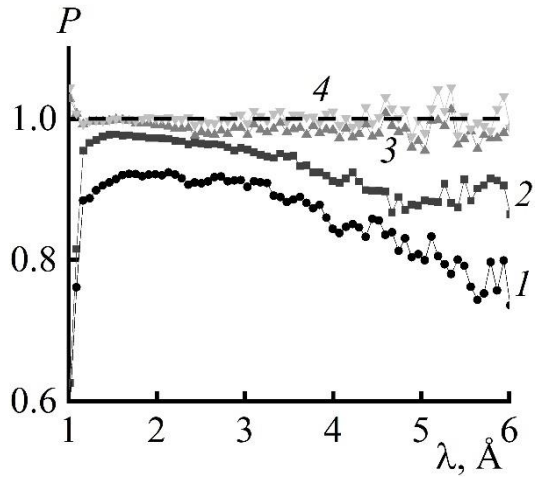

**Fig. 12.** Wavelength dependences of the polarization efficiency of the polarizer $P_p(1)$, of the neutron polarization analyzer $P_a(2)$ and the probability of spin-flip polarization by input (3) and output (4) spin flippers.

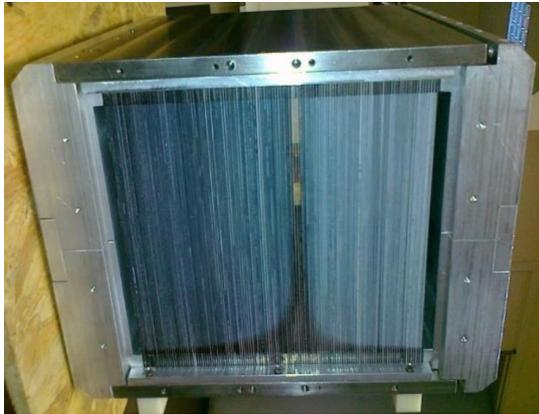

**Fig. 13.** Super-mirror analyzer of neutron polarization of the REMUR reflectometer.

The polarization analyzer (Fig. 13) is performed in the "fan" geometry [18] and manufactured at PNPI (Gatchina). The analyzer is a stack of 125 super-mirrors positioned in a magnetic field of a permanent magnet and has an area of window assumed for neutrons with a cross section of 18×20cm$^2$. Super-mirrors are 0.5 mm-thick glass substrate coated with a m2 type super-mirror coating, characterized by a critical neutron wave vector of 22 mrad/Å (indicated as m2). The mirrors are oriented so as to provide the same glancing angle of the neutron beam on them, equal in this case to 3 mrad. Figure 12 shows that $P_a(\lambda)$ (curve 2) exceeds 0.9 in the wavelength range 1.15 – 4.5 Å.

Gradient radio frequency spin flippers [19, 20] manufactured at PNPI (Gatchina), operating at a frequency of 75 kHz are used in the spectrometer. The probability of neutron beam polarization flip $f_1 = f_2$ in the range $\lambda > 1.15$ Å exceeds 0.99 (Fig. 12, curves 3 and 4). For registering neutrons, a He-3 gas-filled position-sensitive neutron detector is used. The detector has a sensitive area of 20×20 cm$^2$. The spatial resolution is 2.5mm. The registration efficiency of neutrons by a detector exceeds 0.5 at $\lambda > 1$ Å.

The beam intensity on the sample is determined by its divergence, which is installed using the cadmium diaphragm positioned at the output of the polarizer. The maximum solid angle of the polarizer sight from the installation place of the sample is $\Omega_{ps} = 2.0 \times 10^{-5}$ str. The polarized neutron intensity on the sample with a beam divergence in the horizontal plane of neutron reflection $2\Delta\theta = 0.7$ mrad is $2\times 10^4$ cm$^{-2}$s$^{-1}$.

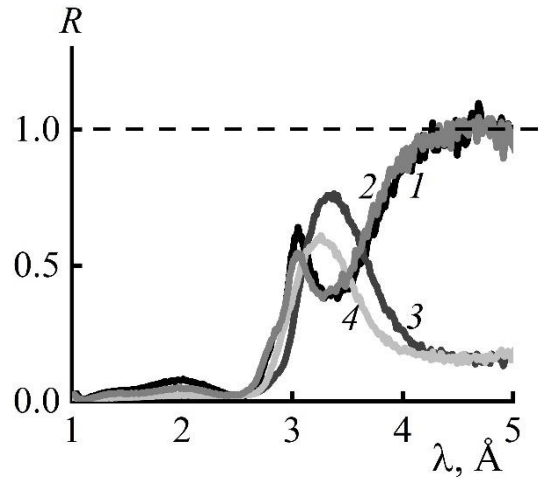

**Fig. 14.** The neutron reflection coefficient from the structure V(10nm)/CoFe(5nm)/$^6$LiF(5nm)/V(5nm)/glass with the pair of (input and output) spin-flippers states "off-off" (1), "on-on"(2), "off-on"(3), "on-off"(4), at the neutron beam glancing angle $\theta=3$ mrad, the magnetic field 295 Oe and the inclination angle of the magnetic field to the plane of the structure 70°.

We now pass over the testing results of the channel. The following structures have been used for testing: V(20 nm)/CoFe(4 nm)/$^6$LiF(5 nm)/V(5 nm)/glass (structure 1), V(20 nm)/CoFe(4 nm)/$^6$LiF(5 nm)/V(15 nm)/glass (structure 2), Cu(10nm)/V(65 nm)/CoFe(4 nm)/$^6$LiF5 nm)/V(5,15 nm)/glass (structure 3) and Cu(10 nm)/V(55 nm)/CoFe(4 nm)/$^6$LiF(5 nm)/V(15 nm)/glass (structure 4) containing the magnetic layer CoFe (5 nm), positioned at a distance of 10 nm (structures 1 and 3) and 20 nm (structures 2 and 4) from a 5 mm-thick glass neutron reflector substrate. Figures 14 and 15 show the reflection coefficient for various spin transitions in the case of the structure 1 (Fig. 14) and the structure 4 (Fig. 15), respectively. As we can see from Fig. 14, the maxima of the dependences with spin flip ("off-on" (3) and "on-off" (4)) corresponds to the minima of the dependences without spin flip ("off-off" (1) and "on-on" (2)), respectively. For the dependences of the reflection coefficient without spin flip for the structure 4 (Fig. 17, curves 1 and 2), two minima are observed in the rage of $\lambda \approx 5$ Å. On the dependence 1, the first minimum corresponds to a single transition "++", the second one - to two consecutive transitions "+ -" and "- +", which in total also provide the transition "++" ("+-" + "-+" ≡ "++"). On the dependence 2, the second minimum corresponds to a single transition "- -", and the first one corresponds to two consecutive transitions "- +" and "+ -", which also provides the transition "- -" ("- -" ≡ "- +" + "+ -").

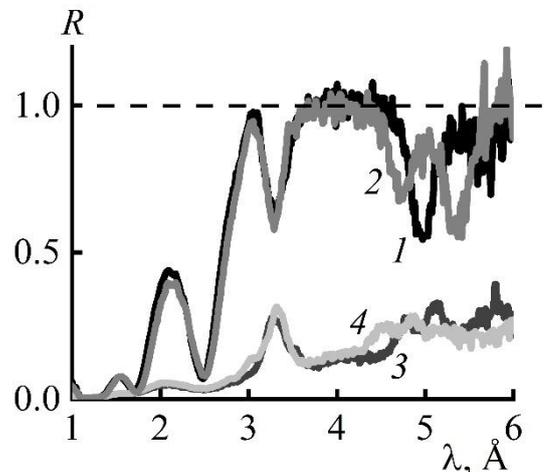

**Fig. 15.** Neutron reflection coefficient from the structure Cu(10 nm)/V(55 nm)/CoFe(5 nm)/6LiF(5 nm)/V(15 nm)/glass of neutrons (1-4) in the state of spin-flippers "off-off"(1), on-on (2), off-on (3), on-off (4), neutron beam

slip angle θ=3 mrad, magnetic field intensity 295 Oe and the magnetic field inclination angle to the plane of the structure 70°.

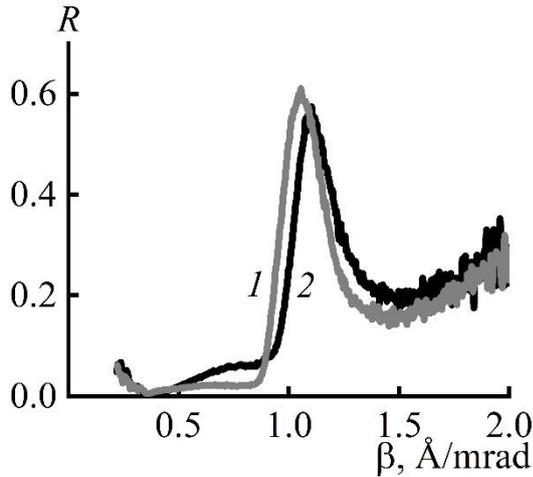

**Fig. 16.** Neutron reflection coefficient from the structures V(20 nm)/CoFe(4 nm)/6LiF (5 nm)/V(5 nm)/glass (1) and V(20 nm)/CoFe(4 nm)/6LiF(5 nm)/V (15 nm)/glass (2) in the state of spin flippers (on-off) at the glancing angle of 3.1 mrad.

Figure 16 shows dependences of the neutron reflection coefficient with spin flip for the structures 1 and 2. From the figure it can be seen that with increasing distance from the magnetic layer to the neutron reflector from 10 to 20 nm, the parameter $\beta_{max} = \lambda/\theta$ for the maximum value of the reflection coefficient, which determines the spatial position of the layer, has been increased by 7%. The experimental value of the uncertainty $\beta$=1.1, related to the uncertainty of the wavelength at $\lambda$=3.1 Å, is 0.65. Hereof it follows that the minimum experimentally determined value of the change in the distance δL due to the uncertainty of λ is 1 nm. The value of the minimum magnetic noncollinearity (perpendicular to the magnetic field of the magnetization value), following from that of equal to the effect of the statistical accuracy of measurements, is also essential. So, at the measurement time of $t = 1$ day, the neutron flux density of $2·10^4 cm^{-2}s^{-1}$ on the sample and the sample cross section of 0.15 $cm^2$, the wavelength of $\lambda$= 3 Å (spectrum maximum at 1.5 Å), the resolution of the neutron wave vector δk/k =0.01 and the thickness of the magnetic layer of 5 nm, for the minimum value of the perpendicular component of the magnetization ΔJ=10 G. For the measurable change of the distance from the substrate to the layer CoFe with a thickness of 5 nm and a perpendicular component magnetization of 2 kG, we have $\Delta L_{flux}$≈0.01nm.

**Channel of charged particles**. The main unit of the channel is the ionization chamber installed in the goniometer of the sample module (Fig. 17). A scheme of the chamber is shown in Fig. 18. A neutron beam enters the chamber through the input window and falls on the structure mounted on the cathode of the chamber. The reflected and refracted on structure neutron beams exit the chamber through the output window and are registered using a position-sensitive detector positioned at a distance of 4.9 m from the chamber. The scattering intensity of neutrons does not exceed 0.5% during their passage through the chamber windows. To adjust the glancing angle of the neutron beam to the plane of the structure, the ionization chamber is oriented by turning around the vertical axis in the range of 1–10 mrad with an accuracy of 0.1 mrad.

The cathode and anode of the ionization chamber with a thickness of 0.5 mm are produced of aluminum. A frame with grooves is fixed on the cathode, where replaceable samples are inserted. The grid is produced from 100 μm -thick gilded tungsten wire, wound in increments of 2 mm on a stainless steel frame. This is positioned in cylindrical stainless steel housing with a length of 300 mm and a diameter of 270 mm, equipped with high-voltage and signal inputs and a valve for pumping and filling the chamber with gas, as well as a control manometer. The neutron beam passes through thin aluminum windows with a thickness of 100 μm and a diameter of 80 mm installed at the ends of the chamber. The distance between the cathode and the grid and between the grid and the collector was 50 and 20 mm, respectively. The sensitive volume of the ionization chamber in the plane of the sample is 210 mm×110 mm and allows the use of samples with a size of up to 170×80 mm. In this study, samples with a size of 71×71 mm containing a layer of the $^6Li$ isotope have been used. A mixture of Ar+3% $CO_2$ at a pressure of 1.5 bar has been used as working gas. At this pressure, the paths of the tritons with an energy of 2.73 MeV from the $^6Li(n,\alpha)^3H$ reaction are in the sensitive volume of the ionization chamber. The ORTEC 142A charge-sensitive preamplifiers are directly installed on the ionization chamber, to which registration signals of a charged particle are supplied from the cathode and anode of the chamber. The ORTEC 660 high-voltage unit and the ORTEC 474 fast amplifiers are positioned in the NIM crate, arranged at a distance of 5 m from the ionization chamber.

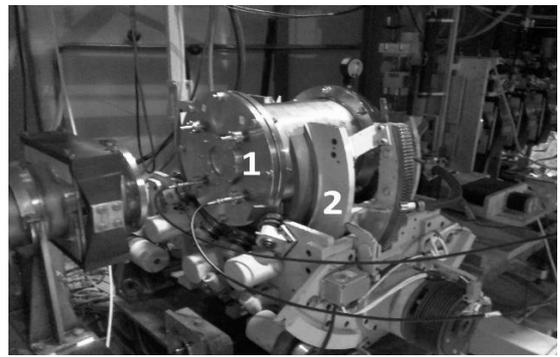

**Fig. 17.** Ionization chamber *1*, installed in the goniometer 2 of the REMUR spectrometer.

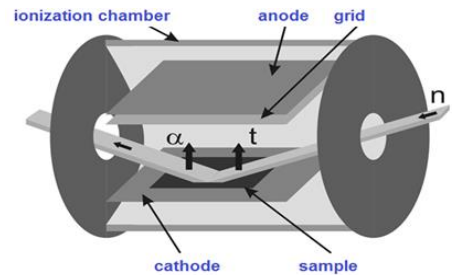

**Fig. 18.** Geometry of the passage of a neutron beam through an ionization chamber.

Depending on the flight time, the intensities of neutrons and charged particles have been registered using neutrons of the distance from the reactor moderator to the detector $t_n$ and the ionization chamber $t_c$, respectively. The distance from the moderator to the neutron detector was $L_n$=34 m and to the ionization chamber – $L_c$=29 m. The experimental dependences of neutron intensities $J_n(t_n)$ and charged particles $J_c(t_c)$ have been converted depending on the neutron wavelength λ, while the following ratio was used:

$$\lambda = \frac{\gamma t_n}{L_n} = \frac{\gamma t_c}{L_c} \quad (10)$$

where $t_n=N_n\Delta t_n$, $t_c = N_c\Delta t_c$, $N_n$, $N_c$ – are the channel numbers of the neutron detector and the ionization chamber, $\Delta t_n$ = 128 μs, $\Delta t_c$ = 100 μs – the width of the channel of the temporary encoder of the neutron detector and of the ionization chamber, respectively, γ = h/m, h is Planck constant, m – neutron mass.

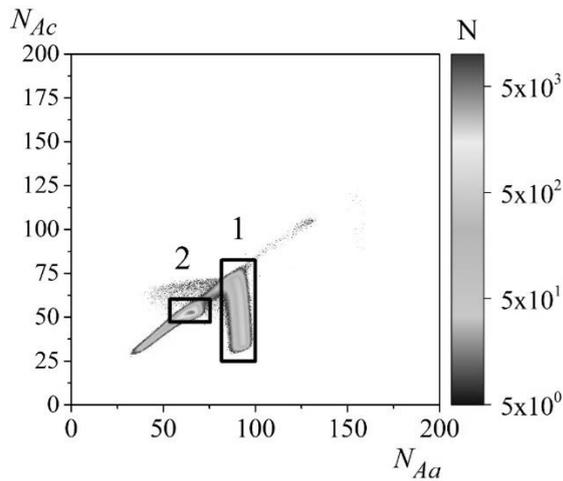

**Fig. 19.** Map of the counting rate distribution of alpha-particles (*1*) and tritons (*2*) in the ionization chamber for the structure No. 4, depending on the amplitudes of signals (channel numbers) from the anode ($N_{Aa}$) and cathode ($N_{Ac}$).

As it follows from (10), the widths of channels are not identical and relate as $\Delta\lambda_n/\Delta\lambda_c = 1.092$. In this regard, $J_c(\lambda)$ was normalized to the width of the neutron channel $\Delta\lambda_n$. For testing the channel of charged particles the same structures have been used as for testing the channel of polarized neutrons.

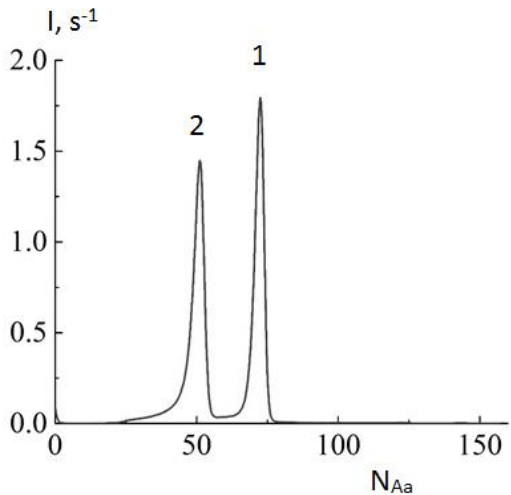

**Fig. 20.** Integrated over signal amplitude from the cathode, dependence of the counting rate of alpha particles (*1*) and tritons (*2*) in the ionization chamber based on the signal amplitude from the anode.

Figure 19 shows the intensity distribution of charged particles on the plane of the signal amplitude from the cathode (ordinate) – signal amplitude from the anode (abscissa). Spot *1* corresponds to $^4$He alpha particles, spot *2* - to $^3$H tritons occurring in the $n+^6Li = {}^4He+{}^3H$ reaction. The amplitude spectrum of registered alpha particles (*1*) and tritons (*2*) is shown in Fig. 20. The intensities of alpha particles and tritons integrated over signal amplitude from the anode are referred to 1: 0.99, that is, both particles are completely absorbed in the chamber.

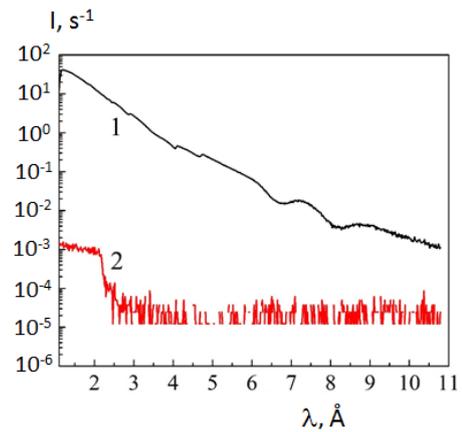

**Fig. 21.** Wavelength dependence (spectrum) of the detector counting rate of neutrons passing through the ionization chamber without interacting with the structure (1) and the detector background counting rate (2).

The wavelength dependences of the neutron counting rate and the background counting rate of the neutron detector are shown in Fig. 21. At the spectrum maximum $\lambda = 1.1$ Å, the neutron counting rate in the channel with a duration of 128 μs ($\Delta\lambda = 0.015$ Å) is 40 s$^{-1}$, and the ratio of the neutron counting rate to the background counting rate is $\eta=4\times10^4$. At $\lambda=2.5$ Å, the background is suppressed by the neutron beam chopper, in this regard, the neutron intensity decreases 5 times, while η increases to $3\cdot10^5$. The background counting rate of the ionization chamber in the open window of the neutron chopper ($\lambda = 1–2$ Å), when the neutron beam is blocked by a cadmium plate, is due to fast neutrons (with energies of several MeV) and gamma radiation and is $10^{-4}$ s$^{-1}$ (the background counting rate of the neutron detector is $10^{-3}$ s$^{-1}$). With the neutron chopper window closed ($\lambda>2.5$ Å), both background counting rate decreases to $4\times10^{-5}$ c$^{-1}$.

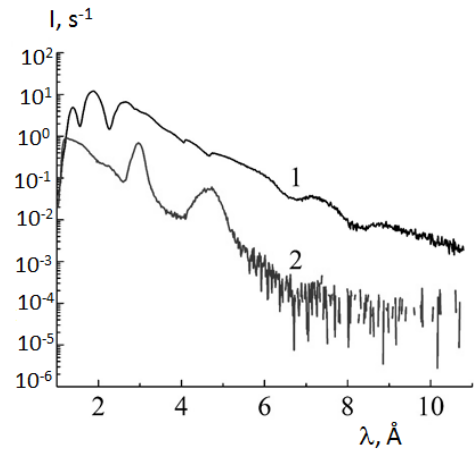

**Fig. 22.** Dependences of the intensity of reflected neutrons (1) and charged particles (2) on the wavelength for the structure No. 4 at a neutron glancing angle of 2.9 mrad (the peak intensity of charged particles at $\lambda = 3.25$ Å in a channel with a duration of 128 μs is 0.52 s$^{-1}$, the intensity, integral over the neutron wavelength, is 20 s$^{-1}$).

The spectra of neutrons and charged particles in the amplified standing wave mode (structure No. 4) are shown in Figure 22. The neutron density gain coefficient is 15 and 11 at wavelengths of 3 and 4.8 Å (curve 2), respectively. These values significantly exceed 4, indicating the implementation of the amplified neutron standing wave regime. It can also be seen that measurements in both channels are possible in the neutron wavelength range of 1–10 Å.

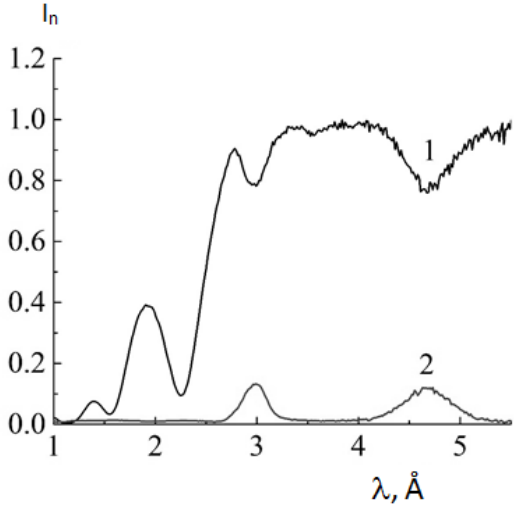

**Fig. 23.** Normalized intensity of reflected neutrons (1) and of charged particles (2) depending on the neutron wavelength at θ=2.85 mrad for the structure No. 4.

The normalized reflected neutron intensity $I_R/I_0 \equiv R$(1) and the normalized charged particle intensity $I_c/I_0 \equiv M_c$(2), where $I_c$ is the intensity and $M_c$ is the coefficient of forming charged particles, relatively, are shown in Fig. 23 for the structure No.4. At λ>3 Å, neutrons are reflected completely, and the relation $R + M_n = 1$ is valid, where $M_n$ is neutron absorption coefficient. When neutrons are reflected completely, the ratio $M_c \leq 0.5 M_n \equiv 0.5\times(1-R)$ should be realized for $M_c$ and $M_n$. In this inequality, firstly, it is considered that the efficiency of registered charged particles does not exceed 50% of the efficiency of registered neutrons due to the fact that those charged particles are registered which fly out only from the cathode to the chamber. Secondly, $M_n$ is determined not only by neutron capture using $^6$Li nuclei, but also by neutron scattering at rough interfaces. In the experiment for λ = 3 and 4.7 Å, $M_c \approx 0.5 M_n$ is observed, that is, scattering at rough interfaces is less as compared to the neutron capture by $^6$Li nuclei.

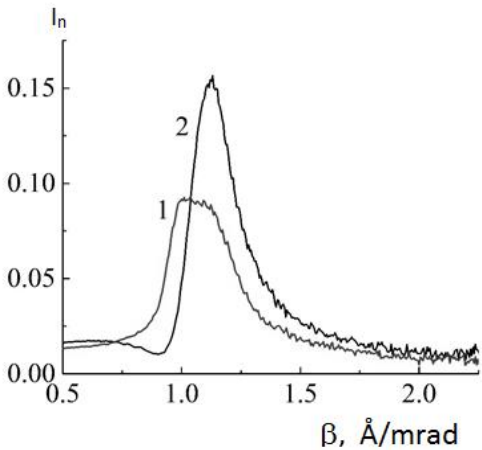

**Fig. 24.** Experimental normalized intensity of charged particles for structures No.1 (1) and No.2 (2).

We now consider the structural information extracted from channels of charged particles. The experimental dependence $I_n(\beta)$, where $\beta = \lambda/\theta$ for the structures No. 1 and No. 2 are shown in Figure 24. Dependence maxima are observed at $\beta_1$=1.085 and $\beta_2$=1.149 Å/mrad, to which $\lambda_1$ = 3.25 and $\lambda_2$ = 3.45 Å correspond. The resulting inequality $\lambda_2 > \lambda_1$ is consistent with the inequality $L_2 > L_1$ for the distance from the layer $^6$Li to the substrate. The experimental data for the initially indicated structures 1-4 are described by the corresponding profiles represented by sets of rectangular potentials. For the structure 1, the profile is 0.5U(5nm)/0.3U(5nm) /-0.044U(10nm) /0.3U(5nm) /0.45U(5nm) /(0.3 - 0.01i) U(1nm)/(0.6-0.02i)U(5nm)/(0.3-0.01i)U(1nm)/-0.044U(9nm)/0.45U(4nm)/0.6U, where U=172 neV is the neutron interaction potential with copper. The thickness of the entire structure 1 is 50 nm, which is 25% more than the value obtained by the method of neutral mass spectrometry.

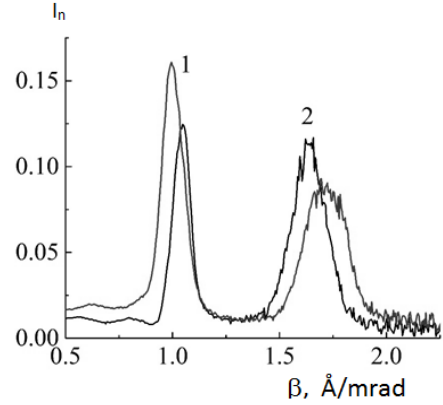

**Fig. 25.** Experimental normalized intensity of charged particles for the structures No 3 (1) and No 4 (2).

For the structure No 2, the profile is 0.9U(5nm)/0.7U(5nm)/-0.044(10nm)/0.3U(5nm)/0.45(5nm)/(0.3-0.01i)U(1nm)/(0.6-0.02i) U(6nm)/(0.3-0.01i)U(1nm)/-0.044U(14.5nm)/0.45U(4nm)/0.6U. The calculated position difference of the $^6$Li layer relative to the substrate for the structures No 2 and No 1 is 5.5 nm, which is less than the nominal value 10 nm. Hence, for a minimum change in the distance of the lithium layer relative to the substrate, resulting from the uncertainty of the wavelength λ, the value δL = 0.55nm follows. Taking into account real profiles for the structures 1 and 2, the similar value δL also follows for the channel of polarized neutrons.

We now estimate the minimum cross section for the neutron interaction with nuclei $\sigma_{min}$ and the minimum change in the distance of the LiF layer from the substrate $\Delta L_{charg}$. Let us assume that δk/k=0.1 for resolution (determined by the beam exit from the polarizer), in this case the neutron flux density on the sample is $2\cdot10^4$cm$^{-2}$s$^{-1}$. Further, we have equation for measurement time $t$

$$I_{eff} t = \sqrt{(2I_b + I_{eff})\cdot t} \qquad (11)$$

where $I_{eff}= I_0\cdot M$, $I_0= J_0\cdot S$, $J_0$ is density of the neutron intensity on the sample, $S$ is the area of the sample cross section perpendicular to the neutron beam, $I_b$ is the background counting rate, $M =\mu\eta$, $\mu=N\sigma h$, $\eta=kn$ /$n_0k_0$, $N$ is the nuclear density, $h$ is the thickness of the neutron-absorbing layer. Next, we take a measurement time of 1 day. Sample is with a cross-sectional area of 0.15 mm, the wavelength is 3 Å, and the maximum in the spectrum is at 1.5 Å. For the measurable cross section of $^6$Li nuclei with a density of $3\cdot10^{22}$cm$^{-3}$ in a 5 nm-thick layer, we have $\sigma_{flux}$=0.2 barn. For a measurable change in the distance from the substrate to the 5 nm-thick LiF layer with a nuclear cross section of 425 barn, we have $\Delta L_{flux} \approx 0.003$ nm.

Figure 25 shows experimental dependences $I_n(\beta)$ for the structures No 3 (curve 1) and No 4 (curve 2). For these structures, an amplified neutron standing wave regime is realized, in which the position of the layer with lithium nuclei is determined by the intensity of charged particles at resonant values of the wave vector. In case of the structure No 3, for which the distance from the lithium layer to the neutron reflector is smaller, the first peak is higher in value, while the second is lower in comparison with the structure No 4. For the structure No 3, the spatial profile is 0.2U(2nm)/1.0U(10nm)/0.2U(2nm)/-0.044U(42nm)/0.3U(5nm)/

0.45U(5nm)/(0.3–0.01i)U(1nm)/(0.6–0.02i)U(4.5nm)/ (0.3-0.01i)U(1nm)/-0.044U(12nm)/0.35U(6 nm)/0.45U(6nm)/0.6U. The total thickness of the structure is 96 nm, which is 7% higher than the thickness value resulting from neutral mass spectrometry data. For the structure No 4, the spatial profile is 0.2U(2nm)/1.0U(10nm)/0.2U(2nm)/- 0.044U(36nm)/0.3U(5nm)/ 0.45U(5nm)/0.3U(1nm)/(0.6-0.02i)U(4 nm)/(0.3 - 0.01i) U(1 nm)/-0.044U(16nm)/0.35U(6nm)/0.45U(6 nm)/0.6U. The total structure thickness is 94 nm, which is 5% higher than neutral mass spectrometry data. The difference between the positions of the $^6$Li layer relative to the substrate for the structures No 3 and No 4 is 4 nm, which is close to the corresponding value for the structures No 1 and No 2.

**Channel of gamma rays.** To register gamma rays, a semiconductor germanium detector operating in the range of 3 keV - 10 MeV was used. The germanium crystal has a diameter of 61.2 mm and a length of 87.3 mm. For gamma radiation with an energy of 1.33 Mev, the registration efficiency is 45%, the energy resolution was 2 keV.

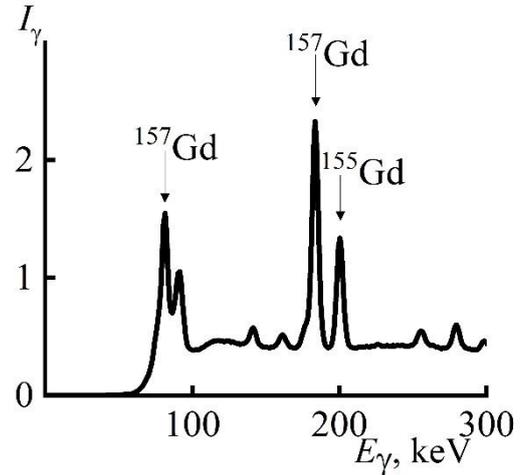

**Fig. 27.** Energy spectrum of gamma rays from the structure No5.

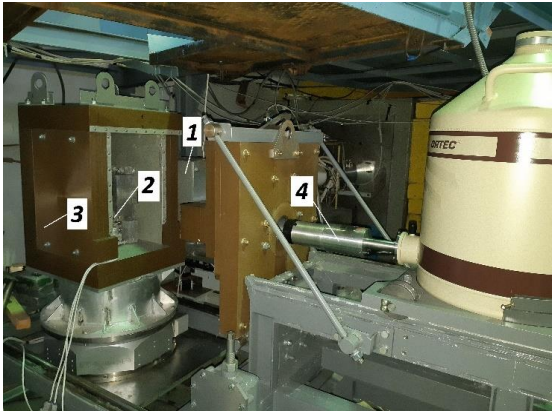

**Fig. 26.** Structural units of the channel for registering gamma radiation: 1 - neutron beam collimator, 2 - sample installation place, 3 - shield for the sample installation place, 4 - gamma detector with cryostat.

Figure 26 shows structural units of the final section of the gamma channel: the neutron beam collimator 1 is produced from borated polyethylene and lead, the installation place of the sample 2 is surrounded by borated polyethylene, the sample is installed on tables providing its movement and rotation (3), the gamma detector 4 is positioned at a certain distance from the sample through a hole in the shield (34 mm without anti-Compton system and 12 cm with it). In the second position, compared to the first one, the signal decreases by 5.3 times, and the background - by 13.7 times. For testing the channel, we used structures containing layers of natural gadolinium, namely, V(20nm)/Gd(5nm)/V(5 nm)/Cu(100nm)/glass(structure No 5), V(10nm)/V(55nm)/Gd(5nm)/V(15nm)/Cu(100nm)/glass(structure No 6), Cu(10 nm)/V(65nm)/Gd(5nm)/V(5nm)/Cu(100nm)(structure No 7) and Cu(10nm)/V(55nm)/Gd(5nm)/V(15nm)/Cu(100 nm)/glass (structure No 8). The strongest transition in the $^{157}$Gd isotope was registered at an energy of 181.94 keV of gamma rays, the ratio of which is $\alpha_{Gd-En}$=0.1833 in the total cross section of the neutron interaction with the isotope. Occurrence of the $^{157}$Gd isotope in the natural mixture is $\alpha_{Gd-Nat}$ = 0.1568. Taking this into consideration, the partial coefficient of the $^{157}$Gd isotope in a natural gadolinium mixture, corresponding to gamma radiation with an energy of 181.94 keV, is $\alpha_P=\alpha_{Gd-En}\times\alpha_{Gd-Nat}$=0.029. Accordingly, the partial cross section of the neutron capture at a neutron wavelength of 1.8 Å is $\sigma_P$=7.3 kbar. As an example, the energy spectrum of gamma radiation from the structure No 5, with neutrons reflecting from it, is shown in Fig. 27.

When registering gamma rays with a certain energy emitted by isotope nuclei, the partial neutron absorption coefficient is determined by an element. We will consider the issue on the experimental determination of the partial neutron absorption coefficient (Fig. 28).

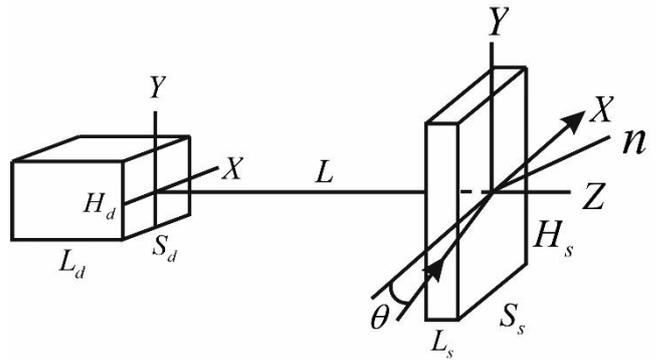

**Fig. 28.** Layout of the test sample, characterized by the geometric parameters $S_s$, $H_s$ and $L_s$ and a gamma detector, characterized by geometric parameters of the registering area $S_d$, $H_d$ and $L_d$.

For the neutron flux absorbed by an isotope in the sample in case of the neutron beam cross section $L_bH_b$ exceeding the sample cross section $S_sH_s\sin(\theta)$, we have

$$J_n = n_0 v_0 S_s H_s \sin(\theta) M \qquad (12)$$

where the absorption coefficient is $M = \int \frac{n(Z_s)k_i^2(Z_s)}{n_0 k_0} dZ_s$

For the flux of gamma rays absorbed in the detector, we have:

$$J_\gamma = S_s H_s \sin(\theta) v_0 M \cdot \alpha_P \cdot$$
$$L^2 \int\int\int\int R^{-4} G(E_\gamma, Y_d, Y_s, X_d, X_s, L) dX_s \cdot dY_s \cdot dX_d \cdot dY_d /$$
$$\int\int dX_s dY_s \qquad (13)$$

where $R=[(Y_d - Y_s)^2+(X_d - X_s)^2 + L^2]^{1/2}$, $G(E_\gamma, Y_d, Y_s, X_d, X_s, L)$ is the registration efficiency of gamma rays with an energy $E_\gamma$ emitted by the surface element $dX_s\, dY_s$ and falling on the surface element of the detector $dX_d\, dY_d$, $M_P = M \times \alpha_P$ is the partial neutron absorption coefficient.

The $X_s$ integration is implemented in the interval $0 \div S_s$ and the $Y_s$ integration - in the interval $0 \div H_s$. In (13), it is assumed that gamma radiation in the investigated sample does not decrease, and the density of falling neutrons is accepted as unity. Calibration of the gamma ray registration channel was carried out based on measurement with a natural cadmium plate $d_{Cd}$ = 1 mm thick, positioned at an angle of $\theta_{Cd}$=25 degrees to the neutron beam (beam glancing angle on the plate). The neutron absorption coefficient in the plate with a bulk density of atoms of natural cadmium is well defined

$$M_{Cd} = (1 - \exp(-N_{Cd}\sigma_{Cd} \cdot \frac{d_{Cd}}{\sin\theta_{Cd}})) \quad (14)$$

The most strongly neutron absorbing isotope with the interaction cross section $\sigma(^{113}Cd, 25$ мэВ$)$=20.6 kbar is $^{113}$Cd. Its content in the natural mixture is 12.26%. The gamma radiation energy of the strongest transition in $^{113}$Cd is $E_\gamma$ = 558.456 keV. The probability of this radiation is $F(558.456$ keV$)$ = 0.74. Thus, for cadmium, $\alpha_P$ = 0.091, and the cross section of the neutron interaction with natural cadmium on the channel with $E_\gamma$ = 558.456 keV in the wavelength range $\lambda$ = 1-10 Å varies from 1.038 to 10.38 kb. Accordingly, at the used values of $d_{Cd}$=1 mm, $\theta_{Cd}$=25° and the tabulated value of the cadmium density $N$(Cd) = 4.6 $\cdot 10^{22}$см$^{-3}$, the neutron absorption coefficient is in the range of $M_{Cd}$ = (1-3.9$\cdot 10^{-6}$) ÷ (1-7.7$\cdot 10^{-55}$), that is, it is practically equal to unit.

In this case, with the beam width $d_n$ (Z axis), which is less than the plate width, we have:

$$J_\gamma = d_n H_s v_0 F(558.456 \text{ кэВ}) \iiint R^{-4} \cdot G(E_\gamma, Y_d, Y_s, X_d, X_s, L)dX_s(^{113}Cd)\, dY_s(^{113}Cd)dX_d dY_d/ \iint dX_s(^{113}Cd)dY_s(^{113}Cd) \quad (15)$$

Here it should be noted that the size $l_x$ of the luminous area (emitting gamma radiation) along the $X_s$ axis slightly exceeds the size of the neutron beam equal to $d_n \times$ctg$(\theta_{Cd})$=2.41 mm. From $M_{Cd}\approx$0.9 it follows that this exceeding is 0.36 mm (15%) at $\lambda$=1 Å and 0.036 mm (1.5%) at $\lambda$=10Å. As a result, $X_s(^{113}Cd)$ integration is carried out in the interval $d_n$ctg$(\theta_{Cd})$+1/$(N(^{113}Cd)\sigma(^{113}Cd))$, and $Y_s(^{113}Cd)$ integration - in the interval of the neutron beam height $H_b$. For low values of the gamma radiation energy (100-300 keV), and also based on the sizes of the registering crystal of the gamma-ray detector, the efficiency $G(E_\gamma, Y_d, Y_s, X_d, X_s, L)$ is equal to one and can be considered with some approximation that does not depend on the coordinates. In this case, for the partial coefficient of neutron absorption by an element in the structure, the relation is

$$M_p = J_\gamma \cdot [F(558.456 keV)L^2(^{113}Cd) \\ \iiint R^{-4} dX_s(^{113}Cd)dY_s(^{113}Cd)dX_d dY_d \iint dX_s dY_s]/ \\ [J_\gamma(^{113}Cd)\sin(\theta)L^2 \iiint R^{-4} \cdot dX_s \cdot dY_s \cdot dX_d \cdot dY_d \iint dX_s(^{113}Cd)dY_s(^{113}Cd)] \quad (16)$$

The spatial profiles of elements for the structures No. 5 and No. 6, obtained by the method of neutral mass-spectrometry (NMS) are shown in Fig. 29 a,b. The wavelength dependences of the neutron reflection coefficient (1, 2) and the normalized partial neutron absorption coefficient $M=M_p/\alpha_P$ in the gadolinium layer (3, 4) for the structures No. 5 and No. 6 are presented in Fig. 30.

It can be seen from Fig. 30 that, at $\lambda_1$=3.1 and $\lambda_2$=3.3 Å, for the corresponding structures, there are minima for the neutron dependences 1 and 2 and maxima for the gamma dependencies 3 and 4 concerning the neutron capture by the nuclei of the $^{157}$Gd isotope. In the range of 3-10 Å the sum is $(R+M) < 1$. This indicates the occurrence of other radiations besides the registered secondary radiation. This additional radiation can be, for example, neutrons scattered in homogeneities. Calculations show that the dependencies in Fig. 30 correspond to the structures described using the formulae A($k_v^2$=42$\cdot$10$^{-6}$ Å$^{-2}$,$d$=10 nm)/[0.35Gd+0.6V](5nm)/[0.6Gd+0.4V](3 nm)/[0.35Gd+0.65V](5 nm)/V(6 nm)/Cu and A($k_v^2$=42$\cdot$10$^{-6}$ Å$^{-2}$, $d$=10 nm)/[0.35Gd+0.6V](5 nm)/[0.6Gd+0.4V](3 nm)/[0.35Gd+0.65V](5 nm)/V(10 nm)/Cu. The effective thickness of the gadolinium layer corresponding to a layer with rectangular spatial distribution and tabulated atomic density was 5.3±0.4 nm, which is close to the originally specified value of 5 nm. Next, the vanadium layer between the copper and gadolinium layers has a thickness of 6 nm for the structure No 5 and 10 nm for the structure No 6, which is at variance with the originally specified values of 10 nm and 20 nm, respectively. The layer A ($k_v^2$=42$\cdot$10$^{-6}$Å$^{-2}$, $d$=10 nm) is an oxide or a mixture of vanadium oxides. For the vanadium oxides VO, V$_2$O$_5$, V$_2$O$_3$ and VO$_2$, $k_v^2$ is, respectively, 34.8$\cdot$10$^{-6}$ Å$^{-2}$, 38.9×10$^{-6}$ Å$^{-2}$, 40.3×10$^{-6}$Å$^{-2}$ and 43.8×10$^{-6}$Å$^{-2}$. From this it follows that in the layer A ($k_v^2$=42×10$^{-6}$Å$^{-2}$, $d$=10 nm), VO$_2$ oxide predominates. For gadolinium, the interaction potential depends on the neutron wavelength. For natural gadolinium, it is determined mainly by the $^{155}$Gd and $^{157}$Gd isotopes. From the data [21, 22], the potential of natural gadolinium is determined as:

$$k_v^2 - ik_w^2 = 3.78 \cdot 10^{-6} \cdot \{7.25 + 5.25 \cdot 10^2 \\ \cdot \frac{[(\frac{81}{\lambda^2} - 26.8) - 54i]}{[4(\frac{81}{\lambda^2} - 26.8)^2 + 108^2]} + 2.36 \cdot 10^3 \\ \cdot \frac{[(\frac{81}{\lambda^2} - 31.4) - 53i]}{[4(\frac{81}{\lambda^2} - 31.4)^2 + 106^2]}\}$$

Here, the neutron wavelength is presented in angstroms, and the squares of the wave vector components - in squared inverse angstroms.

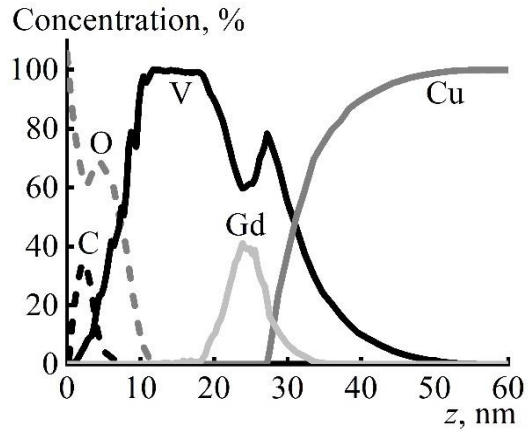

**Fig. 29a.** Spatial profiles of elements for the structure No 5 - V(20 nm)/Gd(5 nm)/V(5 nm)/Cu(100 nm)/glass.

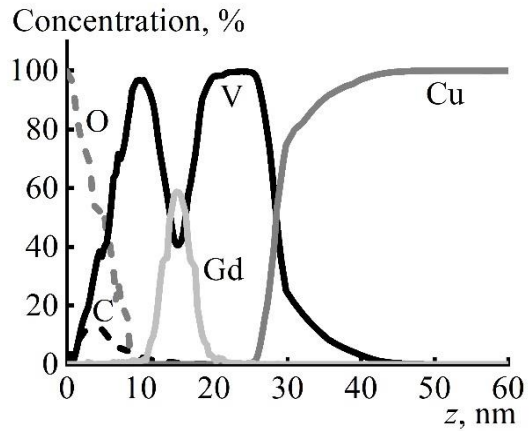

**Fig. 29b.** Spatial profiles of elements for the structure No 6 - V(10 nm)/Gd(5 nm)/V(15 nm)/Cu (100 nm)/glass.

From the dependences presented in Fig. 29 it follows, that the difference between minima depending on the neutron reflection coefficient or maxima depending on the gamma radiation intensity for two structures is $\Delta\lambda=\lambda_2-\lambda_1=0.18$ Å ($\lambda=3.24$ Å) and this corresponds to a change in the distance between the gadolinium layer and the copper reflector equal to $\Delta z = 4$ nm. Considering that the resolution along the wavelength is $\delta\lambda= 0.02$ Å, we obtain the minimum value $\Delta L_{Res}=\delta\lambda\times\Delta z/\Delta\lambda \approx 0.45$ nm that follows from the resolution.

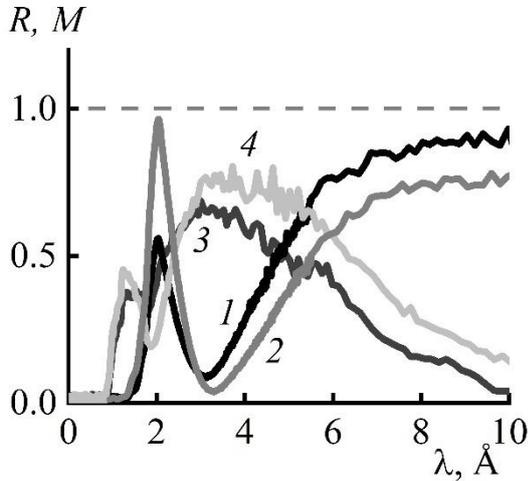

**Fig. 30.** Wavelength dependences of the neutron reflection coefficient (1, 2) and partial neutron absorption coefficients (3, 4) for the structures No 5 (1, 3) and No 6 (2, 4).

Let us now evaluate the minimal values $\sigma_{flux}$ and $\Delta L_{flux}$ for the gamma channel, which follow from gamma flux. The sample has a cross sectional area of 0.075 mm, a wavelength of 3 Å and a maximum in the spectrum at 1.5 Å. With a measurement time of 1 day for a measurable minimum nuclear cross section with a density of $3\cdot 10^{22}$ cm$^{-3}$ in a 5 nm-thick layer, we have $\sigma_{flux} = 4$ barn. For the minimum measurable change of the distance from the substrate to the layer of 5 nm-thick natural gadolinium (registration of only the line $E_\gamma= 181.94$ keV) with a density of $3\cdot 10^{22}$ cm$^{-3}$ and nuclear cross section of 7.3 kbar, we have $\Delta L_{flux}\approx 0.003$ nm.

The spatial profiles of elements for the structures No 7 and No 8 are shown in Fig. 33.

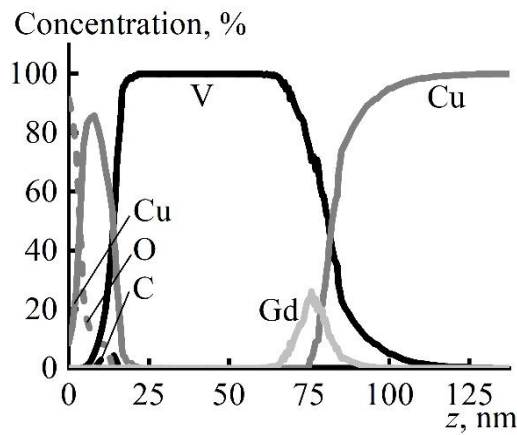

**Fig. 31a.** Spatial profiles of elements for the structure No 7 - Cu(10 nm)/V(65 nm)/Gd (5 nm)/V(5 nm)/Cu(100 nm)/glass.

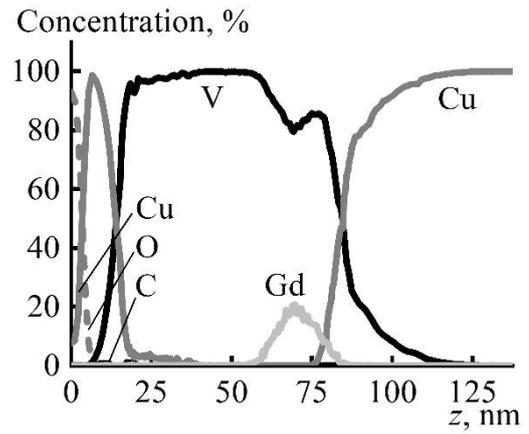

**Fig. 31b.** Spatial profiles of elements for the structure No 8 - Cu(10 nm)/V(55 nm)/Gd(5 nm)/V(15 nm)/Cu(100 nm)/glass.

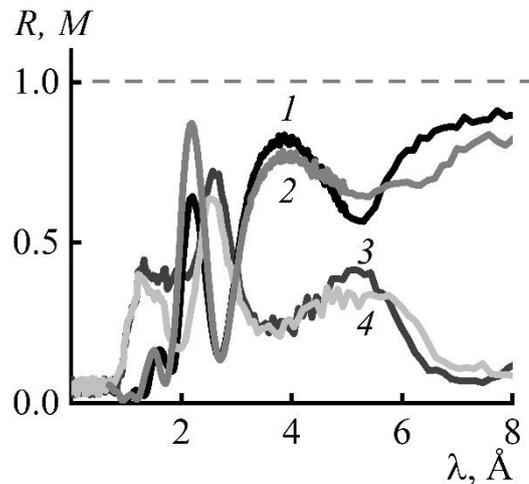

**Fig. 32.** Wavelength dependences of the neutron reflection coefficient (1, 2) and normalized partial neutron absorption coefficients (3, 4) for the structures No 7 (1, 3) and No 8 (2, 4).

The wavelength dependences of the neutron reflection coefficient (1, 2) and normalized partial neutron absorption coefficients (3,4) for the structures No 7 and No 8 are shown in Fig. 32. In the range of total neutron reflection ($\lambda>2$Å) from the copper layer, there are two minima in the neutron dependences and two corresponding maxima in the gamma dependences at resonant values of the neutron wavelength $\lambda = 2.67$ and $5.24$ Å. From these dependences follows the structure Cu(12nm)/[0.7Cu+0.3V](6 nm)/V(43.5 nm)/[0.85V+0.15Gd](5.5 nm)/[0.08V+0.92Gd](4 nm)/[0.85V+0.15Gd] (5.5nm)/[0.1Cu+0.9V](1nm)/ [0.7Cu+0.2]V(2nm)/Cu and Cu(12nm)/[0.7Cu+0.3V](6nm)/ V(40.5nm)/[0.85V+0.15Gd] (5.5nm)/ [0.08V+0.92Gd](4 nm)/ [0.85V+0.15Gd](5.5 nm)/[0.1Cu+0.9V](8nm)/[0.7Cu+0.2V](2 nm)/Cu. The effective thickness of the gadolinium layer was 5.3 nm. The change of the peak amplitude in the gamma channel at $\lambda=2.67$ and at $\lambda=5.24$ is 7% (the statistical error in one channel is less than 0.7%) and 20% (the statistical error in one channel is 5%), respectively.

It is crucial to register gamma rays from magnetic elements. When this occurs, in the case of non-collinear magnetic structure it is possible to register both types of secondary radiations (neutrons with spin flip and gamma radiation). The latter increases the accuracy of determining the magnetic structure and its components of the spatial profiles. However, only gadolinium (Gd) and dysprosium (Dy) have high enough neutron interaction cross-section. The other rare-earth (Tb, Ho, Er, Tm) and transitional elements (Fe, Co, Ni) have relatively low cross sections. However, an attempt was made to observe gamma rays from the cobalt layer.

Natural cobalt wholly consists of the $^{59}$Co isotope and for thermal neutrons ($\lambda$=1.8 Å) it has interaction cross section of 37.2 barn. The most powerful gamma-ray line has the energy 229.72 Kev, its intensity is 15.18% (partial coefficient is $\alpha_p$=0.1518) relative to the full intensity of the gamma radiation. As a result, this gamma line corresponds to the neutron interaction cross section $\sigma_p$=5.65 barn. The first measurements were performed with the Co(60 nm)/glass structure. The area of the glass substrate was 40 cm$^2$. At the gamma-ray energy $E_\gamma$=230 keV, a peak is observed, amounting to 0.06 with respect to the background intensity. The normalized partial neutron absorption coefficient for the Co(60 nm)/glass structure (curve 1) is shown in Fig. 33. At a wavelength of $\lambda$=1.9 Å, the absorption coefficient maximum in the cobalt layer is observed, relating to an increase of neutron density at the substrate critical wavelength $\lambda_{sub}$=1.9 Å.

Measurements were also carried out with the periodic structure Nb(10nm)/10×[Co(12.5 nm)/Si(12.5 nm)/Al$_2$O$_3$ (Fig. 33, curve 2)). The absorption coefficient maximum is observed at a wavelength of 2.4 Å. This wavelength is the average between the Bragg wavelength $\lambda_B$=1.5 Å for the periodic structure and the cobalt critical wavelength $\lambda_{Co}$=3.4 Å.

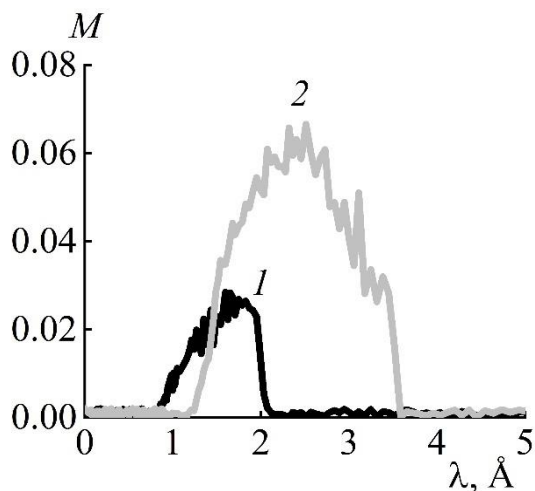

**Fig. 33.** Wavelength dependence of the partial neutron absorption coefficient in the cobalt layer for the structure Co(60 nm)/glass (1) at $\theta$=1.7 mrad and $S$=10 cm$^2$, as well as in cobalt layers for the structure Nb(10 nm)/10×[Co(12.5 nm)/Si(12.5 nm)/Al$_2$O$_3$(2) at $\theta$=3.06 mrad and $S$=1cm$^2$.

**Conclusion.** Currently, several dozen isotopes and magnetic elements are already available for measurements on the spectrometer REMUR. Further development here refers to the following stages. The first stage is the increase based on the use of neutron intensity of neutron guide on the structure in 5 - 10 times. The second is the decrease due to more efficient safety of the fast neutron background and gamma radiation from the reactor core in 5-10 times. The third is the increase of the solid angle of the gamma-ray detector visibility in 4 times or an increase in the number of gamma detectors to 4. Simultaneous implementation of these capabilities will make it possible to increase the neutron interaction cross section with a nucleus to 1 mbarn by the use of a 5 nm-thick absorbing layer or decrease the thickness of the investigated layer to 1 Å when a cross-section is 50 mbarn. The spatial resolution in the structures with single layers or bilayers can be reduced to 1 Å when applying super-specular neutron reflector with $k_v \approx 0.09$ Å$^{-1}$[23] in the structure. In the case of studying periodic structures, the atomic spatial resolution can be achieved by decreasing the period of the structure. At this technological level, obviously, values of the period 1 nm are achievable, that provides for the spatial resolution 1-2 Å.